\documentclass[12pt]{article}

\usepackage{epsfig}

\def\be{\begin{equation}}
\def\ee{\end{equation}}
\def\ba{\begin{array}{c}}
\def\ea{\end{array}}
\def\p{\partial}
\def\ben{$$}

\def\een{$$}
\begin{document}

\titlepage

  \begin{center}{\Large \bf
Imaginary cubic oscillator \\
and its square-well approximations \\
in $x-$ and $p-$representation

  }\end{center}

\vspace{5mm}

  \begin{center}
Miloslav Znojil

 \vspace{3mm}

Theory Group, Nuclear Physics Institute ASCR,\\
250 68 \v{R}e\v{z}, Czech Republic

{e-mail: znojil@ujf.cas.cz }

\end{center}

\vspace{5mm}

\section*{Abstract}

Schr\"{o}dinger equation with  imaginary ${\cal PT}$ symmetric
potential $V^{}(x) = i\,x^3$ is studied using the numerical
discretization methods in both the coordinate and momentum
representations. In the former case our results confirm that the
model generates an infinite number of bound states with real
energies. In the latter case the differential equation is of the
third order and a square-well, solvable approximation of kinetic
energy is recommended and discussed. One finds that in the
strong-coupling limit, the exact ${\cal PT}$ symmetric solutions
converge to their Hermitian predecessors.

 \vspace{9mm}

\noindent
 PACS 03.65.Ge

\vspace{29mm}

\subsection*{Acknowledgements}

Work supported by the GA\v{C}R grant Nr. 202/07/1307,  by the
M\v{S}MT ``Doppler Institute" project Nr. LC06002 and  by the
Institutional Research Plan AV0Z10480505.


\newpage

\section{Introduction}

An interest in imaginary cubic anharmonic oscillators dates back to
their perturbation analysis by Caliceti et al \cite{Caliceti}. The
simplified homework example with the mere two-term non-Hermitian
Hamiltonian
 \ben
 H_{BZ}=p^2+i\,x^3
  \een
has been proposed by D. Bessis and J. Zinn-Justin who had in mind
its possible applicability in the context of statistical physics
\cite{DB}. The example has been re-vitalized by C. Bender et al due
to its possible methodical relevance in relativistic quantum field
theory \cite{BM}. They emphasized that the apparent reality of the
spectrum of energies $E_{BZ}$ is quite puzzling.

The conjecture of a full compatibility of similar Hamiltonians with
the postulates of quantum mechanics  \cite{BBjmp} opened many new
and interesting questions. The current Hermiticity of Hamiltonians
was replaced by a weaker condition of their commutativity with a
product ${\cal PT}$ of the spatial parity ${\cal P}$ and of the
complex conjugation ${\cal T}$. The latter factor is to be
understood as a one-dimensional version of the operator of time
reversal.

The discussion between C. Bender and A. Mezincescu \cite{Mezincescu}
pointed out that one of the key problems of the new studies lies in
the ambiguity of the spectrum which depends quite crucially on our
choice of the boundary conditions which can be, in general,
complexified \cite{BT}. The fragile character of the reality of the
energies has been confirmed by the WKB and perturbative studies
\cite{Alvarez,Pham} and by the quasi-exact and exact models
\cite{QES} where the admissible unavoided level crossings
\cite{ptho} prove sometimes followed by the spontaneous breakdown of
${\cal PT}$ symmetry \cite{quartic}.

In such a context we propose here an extremely elementary approach
to similar models replacing interactions which admit just a
numerical treatment (typically, $V(x)=ix^3$) by their exactly
solvable square-well analogues.

\section{Models in coordinate representation}

In a search for analogies between the solvable and unsolvable models
in one dimension, all of the possible forms of a confining well are
often being approximated by the ordinary real and symmetric square
well
  \be
 V^{(SQW)}(x) =
 \left \{
 \begin{array}{ll}
  S^2,& x \in (-\infty,-\pi)\bigcup (\pi,\infty),\\
 0,
 &
   x \in (-\pi,\pi).
   \ea
   \right .
 \label{SQW}
  \ee
In this spirit one can also replace the antisymmetric and imaginary
homework potential $V_{BZ}(x)=i\,x^3 $ by its elementary square-well
analogue
  \ben
 V^{(ISQW)}(x) =
 \left \{
 \begin{array}{ll}
  -i\,T^2,& x \in  (-\infty,-\pi),\\
 0,  &    x \in (-\pi,\pi),\\
  +i\,T^2,& x \in  (\pi,\infty).
   \ea
   \right .
  \een
Schr\"{o}dinger equation which appears in such a setting,
 \be
\left [ -\frac{\hbar}{2m}\,\frac{d}{dx^2}+V^{(ISQW)}(x) \right ]
\psi(x)=E\psi(x)
 \label{ISQW}
 \ee
will be complemented   by the standard ${\rm L}^2(l\!\!R)$ boundary
conditions
 \be
\psi(\pm \infty) = 0.
 \label{bc}
 \ee
The well known ${\cal PT}$ symmetric normalization convention will
be employed, with a free real parameter $G$ in the unbroken ${\cal
PT}$-symmetry requirements \cite{norma}
 \be
  \psi(0)=1,
 \ \ \ \ \ \ \ \p_x \psi(0)=i\,G.
  \label{nor}
 \ee
Putting $\hbar=2m=1$ and using the ansatz
  \be
 \psi(x) =
 \left \{
 \begin{array}{llc}
   \cos k\,x + B\,\sin k\, x,
    & x \in (0,\pi),& k^2 = E,\\
 (L + i\,N)\,\exp (-\sigma\,x)
  ,& x \in (\pi,\infty),&
\sigma^2=i\,T^2-k^2,
   \ea
   \right .
 \label{ansatz}
  \ee
we shall guarantee the  full compatibility of such a convention with
the symmetry requirements (\ref{nor}) by the choice of the purely
imaginary constant $B =i\,G/k$ in wave functions (\ref{ansatz}).

\section{Matching conditions at $x=\pi$}

Let us split $\sigma = p + i\,q$ in its real and imaginary part with
a fixed sign,  $p,q \geq 0$. This gives $p^2+k^2=q^2$ and $2pq=T^2$.
These rules are easily re-parameterized in terms of a single
variable $ \alpha$,
 \be
p=q\,\cos \alpha, \ \ \ \ \ k =q\,\sin \alpha, \ \ \ \ \ q
=\frac{T}{\sqrt{2\cos \alpha}}, \ \ \ \ \ \  \alpha \in (0, \pi/2).
 \label{param}
 \ee
The standard matching at the point of discontinuity is immediate,
 \ben
 \cos k\pi + B\, \sin k \pi =
 (L + i\,N)\,\exp (-\sigma\,\pi),
 \een
 \ben
 -\sin k\pi + B\, \cos k \pi =
 -\frac{\sigma}{k}(L + i\,N)\,\exp (-\sigma\,\pi).
 \een
After we abbreviate $\sigma/k=-\tan \Omega\pi $, we get an
elementary complex condition of  matching of logarithmic derivatives
at $x=\pi$,
 \be
 G = -i\,k\,\tan(k+\Omega)\pi.
 \label{secular}
 \ee
The real part defines our first unknown parameter, $G = G(\alpha)$.
Due to our normalization conventions, the imaginary part of the
right-hand-side expression must vanish, ${\rm
Re}[\tan(k+\Omega)\pi]=0$. An elementary re-arrangement of such an
equation acquires the form of an elementary quadratic algebraic
equation for $X=\tan k\pi$. Its two explicit solutions read
 \be
X_1 = \frac{p+q}{k}, \ \ \ \ \ \ \ \ \ \ X_2 = \frac{p-q}{k}
 \label{sice}
 \ee
or, after all the insertions,
 \be
 \tan \left [
{\frac{ \pi T \sin \alpha^{(+)}}{\sqrt{2 \cos \alpha^{(+)}}
 }} \right ]=
 {\rm tan} \left [ \frac{\pi-\alpha^{(+)}}{2} \right ],
 \label{sicep}
 \ee
 \be
 \tan \left [
{\frac{ \pi T \sin \alpha^{(-)}}{\sqrt{2 \cos \alpha^{(-)}}
 }} \right ]=
 \tan \left [- \frac{\alpha^{(-)}}{2} \right ].
 \label{sicem}
 \ee
In implicit manner these equations specify the two respective
infinite series of appropriately bounded real roots
$\alpha=\alpha^{(\pm)}_n \in (0, \pi/2)$.

\section{Energies}

For $ \alpha \in (0, \pi/2)$ the left-hand-side arguments $[ \ldots
]$ in eqs. (\ref{sicep}) and (\ref{sicem}) run from zero to
infinity. Their tangens functions oscillate infinitely many times
from minus infinity to plus infinity. Within the same interval, the
limited variation of the argument $\alpha$ makes both the eligible
right-hand side functions monotonic, very smooth and bounded, $ {\rm
tan} [{(\pi-\alpha^{(+)})}/{2}] \in (1,\infty)$ and $ {\rm tan}[
{\alpha^{(-)}}/{2}] \in (0,1)$. {\it A priori} this indicates that
our roots $k=k(\alpha_n^{(\pm)})$ will all lie within well
determined intervals,
 \ben
 k_n^{(+)} \in
 \left ( n+\frac{1}{4},n+\frac{1}{2} \right ),
 \ \ \ \ \ \ \ \ n = 0, 1, \ldots,
 \een
 \ben
 k_m^{(-)} \in
 \left ( m+\frac{3}{4},m+{1} \right )
 \ \ \ \ \ \ \ \ m = 0, 1, \ldots.
 \een
After such an approximate localization of the roots, an unexpected
additional merit of our parametrization (\ref{param}) manifests
itself in an unambiguous removal of the tangens operators from both
eqs. (\ref{sicep}) and (\ref{sicem}). This gives the following two
relations,
 \ben
 k_n^{(+)}= n+\frac{1}{2}-\frac{\omega_{n}^{(+)}}{4}
 , \ \ \ \ \ \ \ \ \
 k_m^{(-)}= m+{1}-\frac{\omega_m^{(-)}}{4},
 \ \ \ \ \ \ \ \ \ \ \omega_n^{(\pm)}=
 \frac{2\alpha_n^{(\pm)}}{\pi}\  \in \ (0, 1).
 \een
After an elementary change of notation with $\omega_n^{(+)}
=\omega_{2n}$ and $\omega_n^{(-)}=\omega_{2n+1}$, we may finally
combine the latter two rules in the single secular equation
 \be
\sin \left (
 \frac{\pi}{2}\omega_N \right )=
 \frac{2N+2-\omega_N}{4T}\cdot \sqrt{2 \cos \left (
 \frac{\pi}{2}\omega_N \right )}
 \ \ \ \ \ \ \ \ N = 0, 1, \ldots,
 \ \ \ \ \ \ \ \ \
 \label{enep}
 \ee
In a graphical interpretation this equation represents  an
intersection of a tangens-like curve with the infinite family of
parallel lines. This is illustrated in Figure~1. The equation
generates, therefore, an infinite number of the real roots $\omega_N
\in (0,1)$ at all the non-negative integers $N = 0, 1, \ldots$.

\begin{figure}[h]                     
\begin{center}                         
\epsfig{file=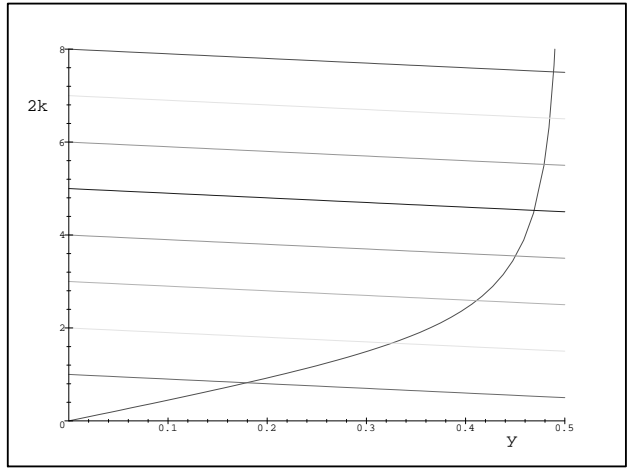,angle=270,width=0.6\textwidth}
\end{center}                         
\vspace{-2mm} \caption{Graphical solution of eq. (\ref{enep}) ($y =
\omega_N/2$, $T=1$).
 \label{obr1a}}
\end{figure}

\section{Wave functions in the weak coupling regime}

Equation (\ref{secular}) in combination with eqs. (\ref{sicep}) and
(\ref{sicem}) determines the real parameter
 \be
G= G^{(\pm)}=-\frac{k^2}{q\pm p}
 \label{ecu}
 \ee
responsible for the behavior of the wave functions near the origin
[remember that $B =iG/k$ in eq. (\ref{ansatz})]. For its analysis
let us introduce an auxiliary linear function of $\omega$ and $N$,\
 \ben
  \sqrt{R(\omega_N,N)}=
 \frac{2N+2-\omega_N}{4T}\  \in \ \left (
 \frac{N+1/2}{2T}\ , \frac{N+1}{2T} \right )\,.
 \een
Our secular eq. (\ref{enep}) can be then read as an algebraic
quadratic equation with the unique positive solution,
 \be
 \cos \left (
 \frac{\pi}{2}\omega_N \right ) = \frac{1}{R(\omega_N,N)
 +\sqrt{R^2(\omega_N,N)+1}} \ .
 \label{secu}
 \ee
This is an amended implicit definition of sequence $\omega_N$. As
long as the right hand side expression is very smooth and never
exceeds one, the latter formula re-verifies that the root $\omega_N$
is always real and bounded as required.

In the domain of the large and almost constant $R \gg 1$ (i.e., for
small square-well heights $T$ or at the higher excitations), our new
secular equation (\ref{secu}) gives a better picture of our
bound-state parameters $\omega_N= 1-\eta_N$ which all lie very close
to one. The estimate
 \ben
 \frac{\pi}{2}\,\eta_N = \arcsin \frac{1}{R+\sqrt{R^2+1}} \approx
 \frac{1}{2R} - \frac{5}{48\,R^3} + \ldots\
 \een
represents  a quickly convergent iterative algorithm for the
efficient numerical evaluation of the roots $\omega_N$. One can
conclude that in a way compatible with our {\it a priori}
expectations, the value of $p=p_N={\rm Re} \sigma \approx q/2R$ lies
very close to zero. As a consequence, the asymptotic decrease of our
wave functions remains slow. We have $q=q_N={\rm Im} \sigma \approx
k$ so that, asymptotically, our wave functions very much resemble
free waves $\exp( -i k x)$. In the light of eq. (\ref{ecu}) we have
$\psi(x) \approx \exp( -i k x)$ near the origin.

\section{Wave functions in the strong coupling regime}

For the models with a very small $R$ (i.e., for the low-lying
excitations in a deep well with $T \gg 1$) we get an alternative
estimate
 \ben
 \frac{\pi}{4}\,\omega_N = \arcsin
 \sqrt{
 \frac{1}{2}
 \left [ R- \left (
 \sqrt{1+R^2}-1
 \right )
 \right ]
 }
  \approx
 \frac{1}{2}\,R-\frac{1}{4}R^2 + \ldots\  \ll  \frac{\pi}{4}.
 \een
In the limit $R \to 0$ the present spectrum of energies moves
towards (and precisely coincides with) the well known levels of
the infinitely deep Hermitian square well of the same width $I =
(-\pi,\pi)$ (cf. eq. (\ref{SQW}) with $S \to \infty$). The
complex-rotation transition from the Hermitian well $V^{(SQW)}(x)$
of eq. (\ref{SQW}) (with $S \gg 1$) to its present non-Hermitian
${\cal PT}$ symmetric alternative $V^{(ISQW)}(x)$ of eq.
(\ref{ISQW}) (with $T \gg 1$) proves amazingly smooth.

Wave functions exhibit a similar tendency. In outer region, they are
proportional to $ \exp ( -px)$ and decay very quickly since $p =
{\cal O}(R^{-1/2})$. Parameter $G^{(\pm)}$ becomes strongly
superscript-dependent,
 \ben
 G^{(+)} = -\frac{k^2}{q+p} ={\cal O}(R^{3/2}),
 \ \ \ \ \ \
 G^{(-)} = -(q+p)={\cal O}(R^{-1/2}).
 \een
In the interior domain of $x \in (-\pi,\pi)$ the wave functions with
superscripts $^{(+)}$ and $^{(-)}$ become dominated by their
spatially even and odd components $\cos kx$ and $\sin kx$,
respectively. The superscript mimics (or at least keeps the trace
of) the quantum number of the slightly broken spatial parity ${\cal
P}$.

We can summarize that our present ${\cal PT}$ symmetric model is
quite robust. Independently of the coupling $T$ the spectrum is
unbounded from above and remains constrained by inequalities
 \be
  \frac{(N+1/2)^2}{4}\ \leq \
E_N \  \leq \ \frac{ (N+1)^2}{4}\
  .
 \label{ene}
 \ee
The analogy between our exactly solvable square-well model and the
standard or ``paradigmatic" ${\cal PT}$ symmetric Hamiltonian
$H_{BZ}$ appears closer than expected.

%
%
%
%

\section{Transition to the momentum representation}

Let us turn our attention to one-dimensional harmonic oscillator
$H^{(HO)}={p}^2+{x}^2$ which is exactly solvable and which appears
in virtually any textbook on quantum mechanics. In  ${\cal PT}$
symmetric quantum mechanics  a similar guiding role can be and has
been attributed to the non-Hermitian cubic Hamiltonian
$H^{(CO)}=p^2+i\,x^3$ of Bender et al \cite{BBjmp}. We have seen
that a straightforward numerical and semi-classical analysis of
the related Schr\"{o}dinger equation
  \be H^{(CO)}
\,|\psi_n\rangle = E^{(CO)}_n \,|\psi_n\rangle, \ \ \ n = 0, 1,
\ldots\ \label{COO} \label{CO}
  \ee
supports a highly plausible conjecture that the spectrum of energies
is real, discrete and bounded below. The conjectured absence of its
imaginary components is  indicated by the Hilbert-Schmidt analysis
\cite{Mezincescu} and by the perturbation calculations in both the
weak-coupling regime \cite{Calicetibe} or in its strong-coupling,
purely numerically generated re-arrangement \cite{norma}.

Certainly, the problem deserves a change of the traditional
perspective. Let us, therefore, move now to its momentum
representation. This would give the momentum as a mere number, $p
\in (-\infty,\infty)$ while the coordinate $x$ becomes represented
by the differential operator $ \hat{x}= i\,\partial_p$. Equation
(\ref{CO}) then acquires a new form containing the {\em purely
real} differential Schr\"{o}dinger operator $H^{(CO)}-E$ of third
order,
  \ben
 \left [ \frac{d^3}{dp^3} + p^2
\right ] \,\psi(p) = E\,\psi(p)\ , \ \ \ \ \ \ \ \psi(p) \in
L_2(I\!\!R).
  \een
This gives an unusual formulation of our bound-state problem where
the quadratic $p-$dependence of the kinetic term $T(p)=p^2$ does not
seem to make the equation any easier to solve. For this reason we
shall drastically simplify the kinetic term and deduce some
consequences.

\section{Piecewise constant approximate kinetic energy }

In a way proposed by Pr\"{u}fer \cite{Pruef} many wave functions can
be visualized as certain deformations of solutions which correspond
to a locally constant potential,  $\psi(x) \approx c_1\sin [\varrho
(x)] + c_2 \cos[ \varrho(x)]$. In the standard quantum mechanics
such a trick found immediate applications in numerical computations
\cite{Ulehla} while it still admits an easy interpretation via some
traditional Sturm Liouvillean oscillation theorems
\cite{Fluegge,Ince}. Using this idea as a methodical guide let us
now replace the kinetic energy operator $T(p) = p^2$ by the most
elementary square well of a finite depth $Z>0$,
  \ben
  T(p) = \left \{
 \begin{array}{ll}
Z, \ \ \  & p \in (-\infty, -1),\\ 0, & p \in (-1, 1),\\ Z, & p
\in (1, \infty). \ea \right .
  \een
In the bounded range of energies $E < Z$ this splits our toy model
in the two separate differential equations,
  \ben
\left [ \frac{d^3}{dp^3} - 8\,\alpha^3 \right ] \,\psi(p) =0, \ \
 \ \ \ \ \ \ \ \ \ \ \ \ p \in (-1,1),
  \een
 \ben
\left [ \frac{d^3}{dp^3} + 8\, \beta^3 \right ] \,\psi(p) =0, \ \
\ \ \ \ \ \ \ \ \ \ \ \ p \in (-\infty, -1) \bigcup (1, \infty). \
\label{eqube}
  \een
The two auxiliary parameters $\alpha=\alpha(E)>0$ and $ \beta=
\beta(E)>0$ are defined in such a way that $Z = E +8\, \beta^3>E =
8\,\alpha^3$. They appear in the three independent (exponential)
solutions of our equation. Their general superpositions are
complex but they may be given the real, trigonometric
 form. Near the origin we have
  \ben
\psi_0(p) = d\,e^{2\alpha\,p} + f\,e^{-\alpha\,p}
\cos(\tilde{\alpha}\, p+\theta) , \ \ \ \ \ \ p \in (-1,1),
 \ \ \ \ \ \ \ \tilde{\alpha} = \sqrt{3}\,\alpha
  \een
where the symbols $d$, $f$ and $\theta$ stand for the three
undetermined constant parameters. In the right and left asymptotic
regions we obtain the similar formulae. After we omit their
exponentially growing and normalization-violating unphysical
components we get the one-parametric family
 \ben
\psi_+(p) = g\,e^{-2 \beta\,p}
 , \ \ \ \ \ \
p \in (1,\infty).
  \een
The two-parametric left-barrier counterpart of this formula reads
 \ben
\psi_-(p) = c\,e^{ \beta\,p} \cos(\tilde{ \beta}\, p+\eta) , \ \ \
\ \ \ p \in (-\infty,-1),
 \ \ \ \ \ \ \ \tilde{\beta} = \sqrt{3}\,\beta
\ .
  \een
At the right discontinuity $p=1$ we have to guarantee the continuity
of $\psi(p)$, $\partial_p \psi(p)$ and $\partial_p^2 \psi(p)$. This
is equivalent to the three matching conditions
 \be
\ba d\,e^{2\alpha} + f\,e^{-\alpha} \cos(\tilde{\alpha}+\theta) =
g\,e^{-2 \beta} ,\\
 2\alpha d\,e^{2\alpha} - \alpha\,f\,e^{-\alpha}\,
[ \cos(\tilde{\alpha}+\theta) +\sqrt{3}\,
\sin(\tilde{\alpha}+\theta) ]= -2 \beta \,g\,e^{-2 \beta} ,\\
 4\alpha^2 d\,e^{2\alpha} - 2\alpha^2f\,e^{-\alpha}\,
[ \cos(\tilde{\alpha}+\theta) -\sqrt{3}\,
\sin(\tilde{\alpha}+\theta) ] = 4 \beta^2 g\,e^{-2 \beta} .
 \ea
\label{match}
  \ee
The weighted sum of these equations re-scales and interrelates
the unknown coefficients $d$ and $g$ in terms of a new energy
parameter $t=t(E)=
\beta(E)/\alpha(E) > 0$ or rather $R=R(E) = (1-t+t^2)^{-1/2} > 0$,
  \ben d\,e^{2\alpha} \equiv  D(E)
 = \frac{
G(E)}{3\,R^2(E)}, \ \ \ \ \ G(E) \equiv g\,e^{-2 \beta}.
  \een
After we eliminate $G$ from the last two equations (\ref{match})
which are linear in $d$ and $f$ we obtain the elementary formula
which defines the shift $\theta=\theta(E)$,
 \ben
\tan (\tilde{\alpha} +\theta) =\frac{\sqrt{3}}{2/t-1}.
 \een
The trigonometric factors become fixed
up to their common sign $\varepsilon=\pm 1$,
 \ben
\cos(\tilde{\alpha}+\theta) = \varepsilon\,[1-t(E)/2]\,R(E), \ \ \
\ \
\sin(\tilde{\alpha}+\theta) =
\frac{\sqrt{3}}{2}\,
\varepsilon\,t(E)\,R(E).
 \een
The same sign enters our last decoupled definition
 \ben
f\,e^{-\alpha} \equiv  F(E)
 = \frac{ 2[t(E)+1]\,
G(E)}{3\,\varepsilon\,R(E)}.
 \een
Up to an overall normalization (say, $g=1$) in our wave functions
matched at $p=1$, all the free parameters become specified as
functions of the energy.

At the second, $p=-1$ discontinuity we have to satisfy three
matching conditions as well. In terms of abbreviations
 \ben
\ba
L_1=
d\,e^{-2\alpha} + f\,e^{\alpha}
\cos(-\tilde{\alpha}+\theta)
 ,\\
L_2=
 2\alpha d\,e^{-2\alpha} - \alpha\,f\,e^{\alpha}\,
[
\cos(-\tilde{\alpha}+\theta) +\sqrt{3}\,
\sin(-\tilde{\alpha}+\theta)
]
 ,\\
 L_3=
4\alpha^2 d\,e^{-2\alpha} - 2\alpha^2f\,e^{\alpha}\,
[
\cos(-\tilde{\alpha}+\theta) -\sqrt{3}\,
\sin(-\tilde{\alpha}+\theta)
]
\ea
 \een
they read
  \ben
\ba L_1(\alpha,d,f,\theta)= c\,e^{- \beta} \cos(-\tilde{
\beta}+\eta)
 ,\\
L_2(\alpha,d,f,\theta)
=
 \beta\,
c\,e^{- \beta} [
 \cos(-\tilde{ \beta}+\eta)
-\sqrt{3}
 \sin(-\tilde{ \beta}+\eta)
 ,\\
L_3(\alpha,d,f,\theta)
=
-2 \beta^2 c\,e^{- \beta} [
 \cos(-\tilde{ \beta}+\eta)
+\sqrt{3}
 \sin(-\tilde{ \beta}+\eta)
 .
\ea
\label{matchdva}
 \een
They determine the values of $c=c(E)$ and $\eta=\eta(E)$. Their
properly weighted sum gives
 \ben
\sqrt{3} \,F(E)\, \sin(-\tilde{\alpha}+\theta) + (2t-1)\,F(E)\,
\cos(-\tilde{\alpha}+\theta) +2\, \frac{1-t+t^2}{t+1} \,
D(E)\,e^{-6\alpha}=0\,.
 \een
After its slight re-arrangement we arrive at the amazingly
transparent relation
 \be
(1-4\,t+t^2)\,\cos(2\sqrt{3}\alpha)
+\sqrt{3}\,(1-t^2)\,\sin(2\sqrt{3}\alpha)
=
\left (
\frac{1-t+t^2}{1+t}
\,
e^{-3\alpha}
\right)^2.
\label{finrel}
 \ee
It should be read as an implicit definition of  physical energies
$E$.

\begin{figure}[h]                     
\begin{center}                         
\epsfig{file=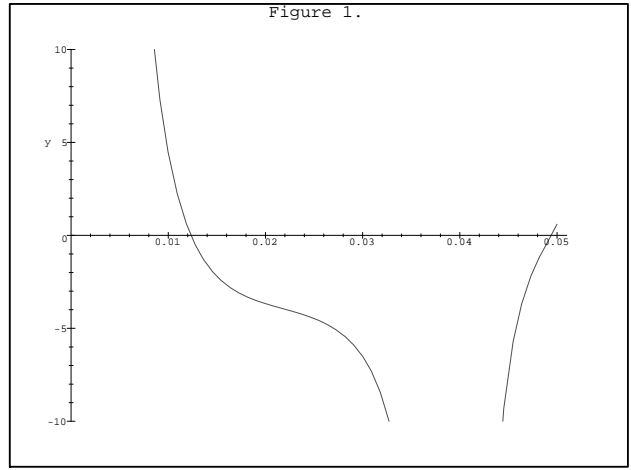,angle=270,width=0.6\textwidth}
\end{center}                         
\vspace{-2mm} \caption{Graphical solution of equation (\ref{finrel})
at $Z = 1/1000$.
 \label{obr2a}}
\end{figure}

\section{The $Z-$dependence of the spectrum}

It is quite instructive to search for the physical energies
numerically. Starting from the very-shallow-well extreme in eq.
(\ref{finrel}) we find the two clearly distinguished energy roots.
The qualitative features of the graph of secular determinant
remain unchanged in a broad interval of the inverse strengths
$1/Z$. Its shape is sampled in Figure 2 at $Z = 10^{-3}$. We have
tested that even the approximate height $\approx -5$ of its left
plateau stays virtually unchanged between $Z = 10^{-5}$ and $Z =
10^{-3}$. Within the same interval of the shallowest wells the
left zero grows from the value $0.00280$ till $0.0241$. Beyond the
broad, downwards-oriented peak one finds the second, right zero
moving from the value $0.01047$ (found very close to the
instantaneous threshold $0.01077$) up the value $0.1056$ (not very
far from its threshold $0.1077$, either) within the same interval
of $1/Z$.

\begin{figure}[h]                     
\begin{center}                         
\epsfig{file=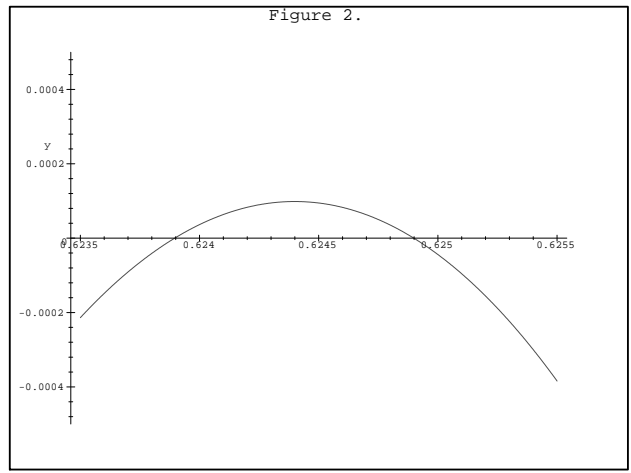,angle=270,width=0.6\textwidth}
\end{center}                         
\vspace{-2mm} \caption{Local maximum giving the new doublet of roots
at $Z = 5.3005$.
 \label{obr3a}}
\end{figure}

A new feature emerges
around $Z = 10^{-1}$ (with the left zero at $0.0445$ and with the
right zero $0.222$ still quite close to the threshold $0.232$) and
$Z = 1$ (with the left zero at $0.072$ and with the right zero
$0.446$, not that close already to the threshold $1/2$)
in the left half of the picture. The plateau develops a local,
safely negative maximum.

In the subsequent domain of $Z > 1$ we have to switch our attention
back to the right half of our graph. Immediately before the coupling
reaches the integer value of $Z=5$, the end of the curve returns to
the negative half-plane near the maximal (i.e., threshold) energy.
This means that there emerges the third energy level there. The
total number of bound states grows to $N=3$ (cf. the leftmost items
in our Table~1).

\subsection*{Table 1. \\
$\mbox{}\ \ \ $ Number of levels $N$ and its changes $\Delta $ with
growing $Z$.}

$$
 \begin{array}
{||c|| ccccccccccccccccccccccc||}
  \hline \hline
 N&2&&3&&5&&3&&4&&5&&7&&8&&6&&7&&9&&10 \\
 \hline
 \Delta&& 1&&2&&-2&&1&&1&&2&&1&&-2&&1&&2&&1&\\
 \hline \hline
  \ea
  $$

\subsection*{}

Beyond $Z=5$, our attention has to return quickly to the left half
of the picture where the very slow growth of the local maximum
creates a new quality at last. The top of the local bump touches and
crosses the horizontal axis at $Z \approx 5.3003$ and $E \approx
0.6244$. At $Z = 5.3005$ a new doublet of energies is formed in a
way illustrated in Figure~3. The number of levels jumps to $N=5$.

\begin{figure}[h]                     
\begin{center}                         
\epsfig{file=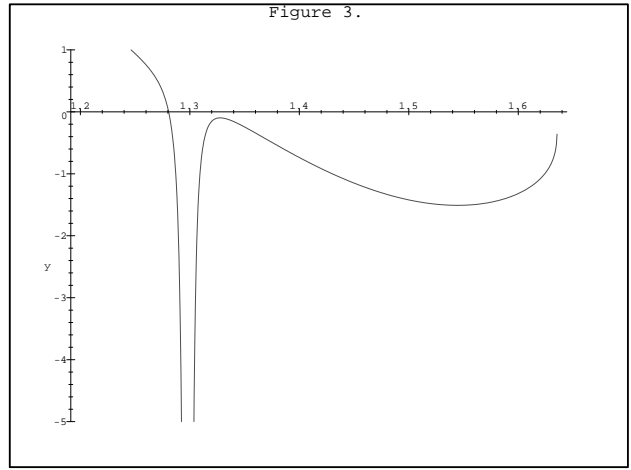,angle=270,width=0.6\textwidth}
\end{center}                         
\vspace{-2mm} \caption{The local maximum not giving the doublet of
roots at $Z = 35$.
 \label{obr4a}}
\end{figure}

A smooth deformation of the graph takes place when the value of $Z$
grows on. During this evolution we discover that our (originally
broad), downward-oriented peak shrinks quite quickly and moves
comparatively slowly to the right.  It gets close to the rightmost
and, to its bad luck, slightly more slowly moving zero number five.
The magnified picture of the resulting ``collision" is displayed
here in Figure 4. At $Z = 35$ it shows that in the threshold region
of the energies,

\begin{itemize}

\item
the wavy motion of the threshold end of our graph still
did not manage to reach the zero axis;

\item
the downwards-oriented peak has already left the positive part
(and moved to the negative part) of the curve in question.

\end{itemize}

 \noindent
As a consequence, the number of levels drops, quite unexpectedly,
down to $3$ again (cf. Table~1).

\begin{figure}[t]                     
\begin{center}                         
\epsfig{file=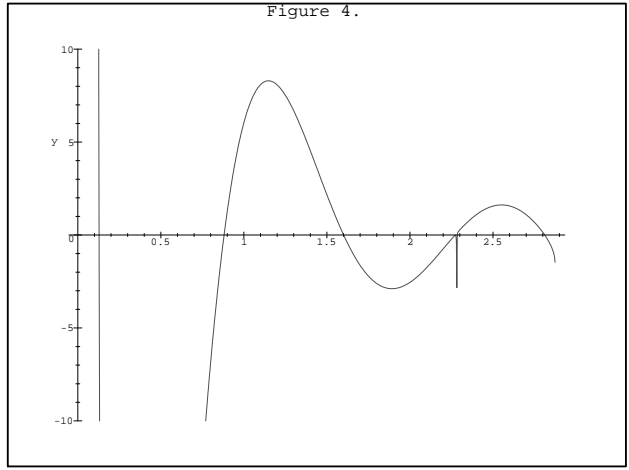,angle=270,width=0.6\textwidth}
\end{center}                         
\vspace{-2mm} \caption{Typical $Z \gg 1$ graph of eq. (\ref{finrel})
($Z = 190$).
 \label{obr5a}}
\end{figure}

In the vicinity of $Z=40$ the new, rightmost energy root emerges at
last. Up to $Z=100$ and beyond, the number of levels stays equal to
$4$. Then it increases to $5$, due to the emergence of the next
threshold zero.  Only after that, the slowly moving downward peak
reaches the domain of the fourth zero. At almost exactly $Z=190$ its
left (and temporarily negative) local maximum reaches the zero value
again (cf. Figure 5). At this moment the number of states jumps up
by two to seven. The magnified graphical proof is offered by Figure
5 at $Z=200$.

\begin{figure}[h]                     
\begin{center}                         
\epsfig{file=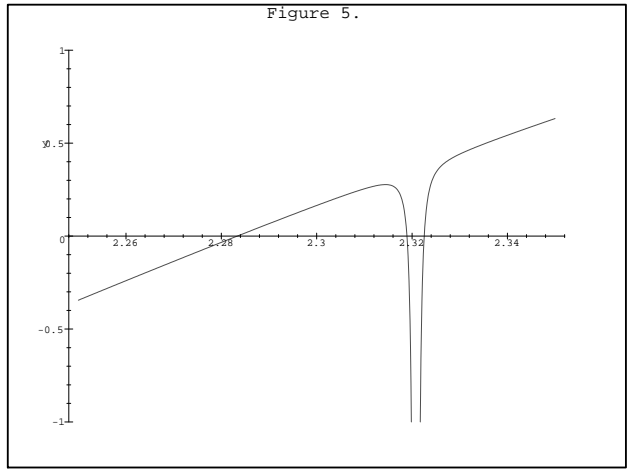,angle=270,width=0.6\textwidth}
\end{center}                         
\vspace{-2mm} \caption{Quasi-degeneracy of the doublet of roots at
$Z = 200$.
 \label{obr6a}}
\end{figure}

The latter Figure illustrates nicely the rapid shrinking of our peak
with $Z$. The numerical detection of its position becomes more and
more difficult. Although this position plays a crucial role in the
practical determination of the number of levels at a given $Z$, we
must be very careful in distinguishing the subgraph of Figure 6
(with three zeros) from a simple straight line with the single zero.

The pattern is deceitful and the standard software which searches
for roots has to be used with due care.  {\it Vice versa}, the above
analysis enables us to take into account all the specific features
of the $Z-$dependence of the graph in eq. (\ref{finrel}). We get a
regular pattern summarized in Table~1 and exhibiting a certain
regularity of the $Z-$dependence of the number of levels $N = N(Z)$.

\section{Energies in the square-well approximation}

The main  consequence of the presence of the above-mentioned narrow
peak is an unusual irregularity observed in the emergence of the new
levels in the deeper wells.  We may conclude that this irregularity
is not an artifact of the computation method. The energy formula
(\ref{finrel}) for our square-well toy model is exact and the
seemingly unpredictable emergence of its roots just reflects the
fact that our Schr\"{o}dinger equation is of the third order. In
particular, there exists no symmetry/antisymmetry with respect to
the parity $p \to -p$ etc.

\begin{figure}[h]                     
\begin{center}                         
\epsfig{file=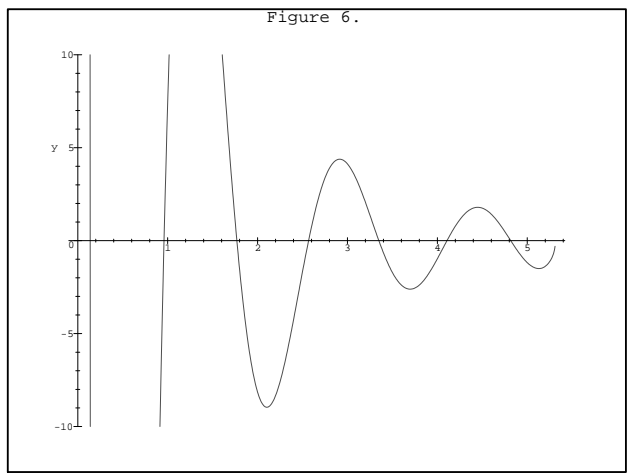,angle=270,width=0.6\textwidth}
\end{center}                         
\vspace{-2mm} \caption{Numerical invisibility of the narrow peak and
of the new threshold root at $Z = 1200$.
 \label{obr7a}}
\end{figure}

Methodical consequences of our analysis are a bit discouraging.
Firstly, the very symbolic-manipulation derivation of our present
formulae proved unexpectedly complicated even in comparison with
the multiple standard square wells in textbooks. That's why we did
not move to any further piece-wise constant approximations
of~$T(p)$.

Secondly, even our use of the most elementary solvable example
revealed quite clearly a very real danger of the possible loss of
certain levels.  For an illustration let us imagine that our
numerical study would have been started in the deep-well domain,
i.e., at the large $Z$.  It is quite easy to generate the graphs
of eq. (\ref{finrel}) there.  In the standard and routine
finite-precision computer arithmetics one discovers that the
results are very smooth and look virtually the same, say, in the
interval of $Z \in (1000, 1200)$.

Let us pick up, for definiteness, the larger sample $Z=1200$.  We
get a picture (cf. our last Figure~7) which is regular and,
deceptively, indicates that $N(1200)=7$.  Unfortunately, the correct
answer (appearing  at the right end of our Table~1) is $N(1200)=10$.
Its derivation requires the use of a significantly enhanced
precision. Otherwise, whenever we use just the standard 14 digits
and Figure~7, we would have missed as much as three (i.e., cca 30 \%
of all) energy levels.

In the light of our preceding considerations, an easy
explanation of the latter numerical paradox lies in the presence
of the narrow peak. {\it A priori}, it is hardly predictable
of course.  It is necessary to spot it by brute force.  One
finds that at $Z=1000$, this anomalous peak still lives safely
below the sixth energy level. The related number of levels is
reliably confirmed as equal to seven, indeed.

In between $Z=1000$ and $Z=1200$, it is necessary to work in an
enhanced precision arithmetics. One finds that the upper, threshold
end of the curve crosses the horizontal axis only slightly above $Z
=1100$.  Due to the very steep slope of the curve in this region,
this crossing is not visible even at $z=1200$ in Figure 7.

One has to trace the narrow peak carefully. It overtakes the sixth
energy level at $Z \approx 1190$ (and $E=4.217$), in an arrangement
resembling our Figure 5 above. Thus, one concludes, finally, that
the new, almost degenerate pair of the energy levels emerges
immediately beyond this point.

\section{Wave functions and their zeros}

A marginal merit of our use of the square-well-shaped $T(p)$ lies in
the availability of the explicit wave functions. For the lack of
space we have to omit  illustrative pictures, mentioning just a few
of their most characteristic features.

In the first step we notice that in the rightmost interval of $p$
the absence of any nodal zero in the wave function is in fact very
similar to the usual Sturm Liouville behavior. Less  expectedly, at
the exact energy value one encounters an infinity of the nodal zeros
in the leftmost subinterval of $p \in (-\infty,-1)$.  In this domain
we are fortunate in studying the exactly solvable case.  The very
presence of this infinite ``left" set of nodal zeros is extremely
sensitive to the numerical level of precision we use. Indeed, the
errors are proportional to the unphysical $\psi^{(unphys.)}(p) \sim
exp(-2 \beta\,p) $ which is growing rapidly at $p \ll -1$.

After the smallest deviation of the energy $E$ from its absolutely
precise bound-state value even the non-numerical and absolutely
precise wave functions will be dominated by the growing asymptotics
$\psi^{(unphys.)}(p) \sim exp(-2 \beta\,p) $ near the left infinity.
The change of sign of the asymptotics is a reliable source of
information about the fact that the energy crossed it physical
value. This observation survives in the shooting numerical
algorithms \cite{Killingbeck} as well as  in the rigorously proved
versions of the  method of Hill determinants \cite{Hill}). In this
context our present numerical experiments could be perceived as
opening a number of new questions. Some of them emerge in the purely
numerical context of an appropriate generalization of the
Pr\"{u}fer-type algorithms. Especially in the vicinity of the
correct physical energies they could lead to reliable and robust
right-to-left shooting numerical recipes.

\section{Outlook}

In the $x-$representation of our problem our main emphasis has been
put on the exact solvability of its replacement by the purely
imaginary square well model. New light has been thrown on some
properties of  wave functions. One can expect that the further
detailed study of the ${\cal PT}$ symmetric square wells will give
new answers to the puzzles concerning the irregular behavior of the
nodal zeros in the complex plane as formulated in ref. \cite{Sturm}.
Our present study indicates that some complexified versions of the
Sturm Liouville oscillation theorems should be developed for the
study of zeros of the separate real and imaginary parts of ${\cal
PT}$ symmetric wave functions.

After the standard Fourier-transformation transition to the
$p-$representation of our imaginary cubic oscillator the underlying
eigenvalue problem can be seen from a different perspective. Its
Hamiltonian is being replaced by a {\em real} differential
expression. On a suitable Hilbert space this specifies the
Hamiltonian operator with the numerical range (and, hence, spectrum)
which is, obviously, real. This complements the extensive discussion
of this topic in \cite{Mezincescu}. Among several immediate {\em
constructive} consequences of the latter observation we underlined
the consistency of the approximations imposed directly upon the
kinetic term $T(p)$.

As long as the behavior of wave-function asymptotics at large $|p|
\gg 1$ differs in the left and right infinity, several new
qualitative aspects of the problem emerge and became clarified by
our schematic piece-wise constant approximation of $T(p)$. At small
$p \approx 0$ the emergence and motion of the nodal zeros can be
interpreted in a graphical manner explaining some features of the
$N(Z)$ dependence. In particular, the puzzling loss of its
monotonicity seems confirmed by our solvable model.

The use of the momentum representation proved able to throw a new
light on the counterintuitive bound states in ${\cal PT}$ symmetric
quantum mechanics.  The emergence/disappearance of our
quasi-degenerate doublets should be emphasized as, perhaps,
analogous to the unavoided level crossings in harmonic oscillators
\cite{ptho} and/or to the anomalous doubling of levels in the models
of Natanzon type \cite{anomaly}. Similar irregularities in the
spectra could be, perhaps, attributed to a peculiar combination of
the analyticity and non-Hermiticity in ${\cal PT}$ symmetric
systems.

In a brief summary of our numerical observations let us point out
the regularity of the $Z-$dependence of the number $N(Z)$ of the
bound states. This indicates that one should search for an improved
application of Sturm-Liouville theory in complex domain
\cite{Hille}. The possibility of  deduction of new oscillation-type
theorems exists, first of all, in the middle interval of $p \in
(-1,1)$ where, for a continuously growing energy parameter $E$, a
steady right-ward movement of the nodal zeros competes with the
exponential terms which are varying slowly.

%
%
%
%

%
%
%
%
%

%
%
%
%
%
%
%
%
%

%
%
%
%

\end{document}